\documentclass[prd,preprint,superscriptaddress,showpacs]{revtex4}
\usepackage{graphicx}

\newcommand{\cdfii}{CDF~II}
\newcommand{\Dzero}{\mbox{\ensuremath{{D^0}}}}
\newcommand{\Dstar}{\mbox{\ensuremath{{D^*}}}}
\newcommand{\jpsi}{\mbox{\ensuremath{{J/\psi}}}}
\newcommand{\dmm}{\mbox{${D^0}\rightarrow\mu^+\mu^-$}}
\newcommand{\dpipi}{\mbox{${D^0}\rightarrow\pi^+\pi^-$}}
\newcommand{\dkpi}{\mbox{${D^0}\rightarrow{K^-}\pi^+$}}
\newcommand{\dstdpi}{\mbox{${D^{*+}}\rightarrow{D^0}\pi^+$}}
\newcommand{\ipb}{\mbox{\ensuremath{\,\mathrm{pb^{-1}}}}}
\newcommand{\mom}{\mbox{\ensuremath{\,\mathrm{GeV/}c}}}
\newcommand{\mass}{\mbox{\ensuremath{\,\mathrm{GeV/}c^2}}}
\newcommand{\dmass}{\mbox{\ensuremath{\,\mathrm{MeV/}c^2}}}

\begin{document}


\begin{center}
{\Large Search for the Flavor-Changing Neutral Current Decay \dmm\ in
$p\overline{p}$ Collisions at $\sqrt{s}$=1.96~TeV}
\end{center}

\font\eightit=cmti8
\def\r#1{\ignorespaces $^{#1}$}
\hfilneg
\begin{sloppypar}
\noindent
D.~Acosta,\r {14} T.~Affolder,\r 7 M.H.~Ahn,\r {25} T.~Akimoto,\r {52}
M.G.~Albrow,\r {13} D.~Ambrose,\r {40}   
D.~Amidei,\r {30} A.~Anastassov,\r {47} K.~Anikeev,\r {29} A.~Annovi,\r {41}
% ...
\end{sloppypar}
\vskip .026in
\begin{center}
(CDF Collaboration)
\end{center}

\vskip .026in
\begin{center}
\r 1  {\eightit Institute of Physics, Academia Sinica, Taipei, Taiwan 11529, 
Republic of China} \\
\r 2  {\eightit Argonne National Laboratory, Argonne, Illinois 60439} \\
\r 3  {\eightit Istituto Nazionale di Fisica Nucleare, University of Bologna,
I-40127 Bologna, Italy} \\
\r 4  {\eightit Brandeis University, Waltham, Massachusetts 02254} \\
\r 5  {\eightit University of California at Davis, Davis, California  95616} \\
\r 6  {\eightit University of California at Los Angeles, Los 
Angeles, California  90024} \\ 
\r 7  {\eightit University of California at Santa Barbara, Santa Barbara, California 
93106} \\ 
\r 8 {\eightit Instituto de Fisica de Cantabria, CSIC-University of Cantabria, 
39005 Santander, Spain} \\
\r 9  {\eightit Carnegie Mellon University, Pittsburgh, Pennsylvania  15213} \\
\r {10} {\eightit Enrico Fermi Institute, University of Chicago, Chicago, 
Illinois 60637} \\
\r {11}  {\eightit Joint Institute for Nuclear Research, RU-141980 Dubna, Russia}
\\
\r {12} {\eightit Duke University, Durham, North Carolina  27708} \\
\r {13} {\eightit Fermi National Accelerator Laboratory, Batavia, Illinois 
60510} \\
\r {14} {\eightit University of Florida, Gainesville, Florida  32611} \\
\r {15} {\eightit Laboratori Nazionali di Frascati, Istituto Nazionale di Fisica
               Nucleare, I-00044 Frascati, Italy} \\
\r {16} {\eightit University of Geneva, CH-1211 Geneva 4, Switzerland} \\
\r {17} {\eightit Glasgow University, Glasgow G12 8QQ, United Kingdom}\\
\r {18} {\eightit Harvard University, Cambridge, Massachusetts 02138} \\
\r {19} {\eightit The Helsinki Group: Helsinki Institute of Physics; and Division
of High Energy Physics, Department of Physical Sciences, University
of Helsinki, FIN-00014 Helsinki, Finland}\\
\r {20} {\eightit Hiroshima University, Higashi-Hiroshima 724, Japan} \\
\r {21} {\eightit University of Illinois, Urbana, Illinois 61801} \\
\r {22} {\eightit The Johns Hopkins University, Baltimore, Maryland 21218} \\
\r {23} {\eightit Institut f\"{u}r Experimentelle Kernphysik, 
Universit\"{a}t Karlsruhe, 76128 Karlsruhe, Germany} \\
\r {24} {\eightit High Energy Accelerator Research Organization (KEK), Tsukuba, 
Ibaraki 305, Japan} \\
\r {25} {\eightit Center for High Energy Physics: Kyungpook National
University, Taegu 702-701; Seoul National University, Seoul 151-742; and
SungKyunKwan University, Suwon 440-746; Korea} \\
\r {26} {\eightit Ernest Orlando Lawrence Berkeley National Laboratory, 
Berkeley, California 94720} \\
\r {27} {\eightit University of Liverpool, Liverpool L69 7ZE, United Kingdom} \\
\r {28} {\eightit University College London, London WC1E 6BT, United Kingdom} \\
\r {29} {\eightit Massachusetts Institute of Technology, Cambridge,
Massachusetts  02139} \\   
\r {30} {\eightit University of Michigan, Ann Arbor, Michigan 48109} \\
\r {31} {\eightit Michigan State University, East Lansing, Michigan  48824} \\
\r {32} {\eightit Institution for Theoretical and Experimental Physics, ITEP,
Moscow 117259, Russia} \\
\r {33} {\eightit University of New Mexico, Albuquerque, New Mexico 87131} \\
\r {34} {\eightit Northwestern University, Evanston, Illinois  60208} \\
\r {35} {\eightit The Ohio State University, Columbus, Ohio  43210} \\  
\r {36} {\eightit Okayama University, Okayama 700-8530, Japan}\\  
\r {37} {\eightit Osaka City University, Osaka 588, Japan} \\
\r {38} {\eightit University of Oxford, Oxford OX1 3RH, United Kingdom} \\
\r {39} {\eightit Universit\'{a} di Padova, Istituto Nazionale di Fisica 
          Nucleare, Sezione di Padova-Trento, I-35131 Padova, Italy} \\
\r {40} {\eightit University of Pennsylvania, Philadelphia, 
        Pennsylvania 19104} \\   
\r {41} {\eightit Istituto Nazionale di Fisica Nucleare, University and Scuola
               Normale Superiore of Pisa, I-56100 Pisa, Italy} \\
\r {42} {\eightit University of Pittsburgh, Pittsburgh, Pennsylvania 15260} \\
\r {43} {\eightit Purdue University, West Lafayette, Indiana 47907} \\
\r {44} {\eightit University of Rochester, Rochester, New York 14627} \\
\r {45} {\eightit The Rockefeller University, New York, New York 10021} \\
\r {46} {\eightit Instituto Nazionale de Fisica Nucleare, Sezione di Roma,
University di Roma I, ``La Sapienza," I-00185 Roma, Italy}\\
\r {47} {\eightit Rutgers University, Piscataway, New Jersey 08855} \\
\r {48} {\eightit Texas A\&M University, College Station, Texas 77843} \\
\r {49} {\eightit Texas Tech University, Lubbock, Texas 79409} \\
\r {50} {\eightit Institute of Particle Physics, University of Toronto, Toronto
M5S 1A7, Canada} \\
\r {51} {\eightit Istituto Nazionale di Fisica Nucleare, Universities of Trieste and 
Udine, Italy} \\
\r {52} {\eightit University of Tsukuba, Tsukuba, Ibaraki 305, Japan} \\
\r {53} {\eightit Tufts University, Medford, Massachusetts 02155} \\
\r {54} {\eightit Waseda University, Tokyo 169, Japan} \\
\r {55} {\eightit University of Wisconsin, Madison, Wisconsin 53706} \\
\r {56} {\eightit Yale University, New Haven, Connecticut 06520} \\
\end{center}
 

\bigskip
\begin{center}
\begin{minipage}{5.6in}
We report on a search for the flavor-changing neutral current decay
\dmm\ in $p\overline{p}$ collisions at $\sqrt{s}$=1.96~TeV using
65\ipb\ of data collected by the CDF~II experiment at the Fermilab
Tevatron Collider.
A displaced-track trigger selects long-lived
\Dzero\ candidates in the \dmm\ search channel, the
kinematically similar \dpipi\ channel used for normalization, the
Cabbibo-favored \dkpi\ channel used to optimize the selection criteria in an
unbiased manner, and their charge conjugates.
Finding no signal events in the \dmm\
search window, we set an upper limit on the branching fraction ${\cal
B}(\dmm)\leq2.5\times10^{-6}$ ($3.3\times10^{-6}$) at the $90\%$
($95\%$) confidence level.

\bigskip
PACS numbers: 13.20.Fc, 14.40.Lb
\end{minipage}
\end{center}
\bigskip

The flavor-changing neutral current (FCNC) decay \dmm~\cite{chconj} is
highly suppressed in the Standard Model (SM) by 
the nearly exact Glashow-Iliopoulos-Maiani (GIM)~\cite{gim} cancellation.
Observation of this
decay at a rate significantly exceeding the SM expectation would
indicate the presence of non-SM particles or couplings.
In the context of the SM, 
Burdman {\it
et al.}~\cite{burdman}\ calculate the branching fraction to be ${\cal
B}(\dmm)\approx10^{-18}$ from short-distance processes, increasing
to ${\cal B}(\dmm)\approx10^{-13}$ when long-distance processes are
included.  This prediction is many orders of magnitude beyond the
reach of the present generation of experiments, whose most stringent
published limits are $4.1\times10^{-6}$ from BEATRICE~\cite{beatrice}
and $4.2\times10^{-6}$ from E771~\cite{e771} at the 90\% confidence
level.  Thus, a large, unexplored region exists in which to search for
new physics.  

Burdman {\it et al.} consider the effects on \dmm\ from a number
of extensions to the Standard Model:
R-parity violating SUSY, multiple Higgs doublets,
extra fermions, extra dimensions, and extended technicolor.
They find that the \dmm\ branching ratio can be enhanced by orders of
magnitude to the range of $10^{-8}$ to
$10^{-10}$ in these scenarios, and in the case of R-parity violating SUSY,
roughly to the level of the existing experimental limit.
Similar enhancements can occur in $K$ and $B$-decays, but charm decays
provide a unique laboratory to search for new physics couplings in the
up-quark sector.


This search uses a 65\ipb\ data sample recorded by the upgraded
Collider Detector at Fermilab (\cdfii) at the Tevatron $p\overline{p}$
collider with $\sqrt{s}=1.96\rm\,TeV$ between February 2002 and January 2003.  
The components of
the \cdfii\ detector pertinent to this analysis are described briefly
below.  Detailed descriptions can be found elsewhere~\cite{cdf}.  CDF
uses a cylindrical coordinate system in which $\phi$ is the azimuthal
angle, $r$ is the radius from the nominal beamline, and $z$ points in
the proton beam direction and is zero at the center of the detector.
The transverse plane is the plane perpendicular to the $z$ axis.  The
pseudorapidity $\eta$ is defined as
$\eta\equiv\tanh^{-1}(\cos\theta)$, where $\theta$ is the polar angle
measured from the $z$ axis.  A silicon microstrip detector
(SVX~II)~\cite{svxii} and a cylindrical drift chamber (COT)~\cite{cot}
immersed in a 1.4~T solenoidal magnetic field track charged particles
in the range $|\eta|<1.0$.  The SVX~II provides up to five $r$-$\phi$
position measurements, each of roughly 15~$\mu$m precision, at radii
between 2.5 and 10.6~cm.  The COT has 96 measurement layers, between
40~cm and 137~cm in radius, organized into alternating axial and
$\pm2^\circ$ stereo superlayers.  The solenoid covers $r<150$~cm,
and electromagnetic and hadronic calorimetry occupy the
region between 150 and 350~cm in radius.  Four layers of planar drift
chambers (CMU)~\cite{cmu} outside the hadron calorimeter cover the
region $|\eta|<0.6$ and detect muons of transverse momentum
$p_T>1.4\mom$ penetrating the 5 absorption lengths of calorimeter
material.

The \Dzero\ decays used in this analysis are selected with a three-level
trigger system.  At the first level, charged tracks are reconstructed
in the COT transverse plane by a hardware processor (XFT)~\cite{xft}.
The trigger requires two oppositely charged tracks with 
reconstructed transverse
momenta $p_T\ge2$\mom\ and $p_{T1}+p_{T2}\ge5.5$\mom.  At the second
level, the Silicon Vertex Tracker (SVT)~\cite{svt} associates SVX~II
position measurements with XFT tracks.  
The impact parameter of the track, $d_0$,
with respect to the beamline, is measured with 50~$\mu$m resolution,
which includes a $\sim30\rm\,\mu m$ contribution from the transverse beam size.
Requiring two tracks with $120\,\mu{\rm m}\le|d_0|\le1.0\rm\,mm$
selects a sample enriched in heavy flavor.
The two trigger tracks must
have an opening angle satisfying $2^\circ\le|\Delta\phi|\le90^\circ$
and be consistent with the decay of a particle traveling
a transverse distance $L_{xy}>200\rm\,\mu m$ from the beamline.
At the third level, a computing farm performs complete event reconstruction.
The sample of $\sim10^5$ \Dstar-tagged two-body \Dzero\ decays selected by
the trigger is used to estimate backgrounds, to optimize selection
requirements, and to normalize the sensitivity of the search from the
data sample itself.

\begin{figure}[htb]
  \begin{center}
    \includegraphics[width=\columnwidth]{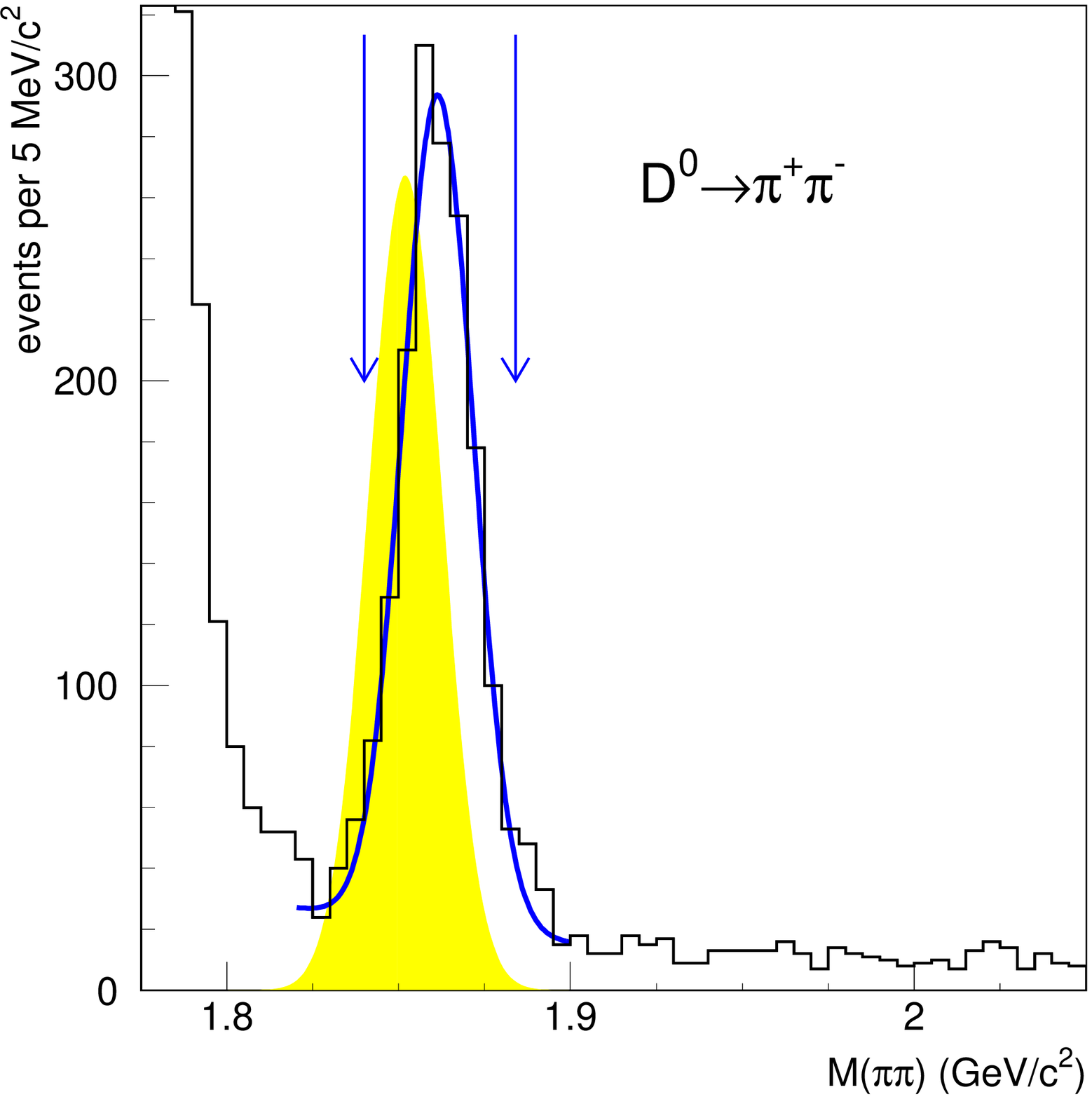}
    \caption{The mass distribution of candidate \dpipi\ events.
    The \dmm\ branching ratio was normalized to the
      kinematically similar mode \dpipi.
      The arrows indicate the $\pm22\dmass$ mass window used for the
        signal.
      The curve is a fit over the range $1.82$ to $1.90{\rm\,GeV}/c^2$ with
Gaussian signal plus linear background.
      The shaded Gaussian represents the effect of reconstructing the
        events with a $\mu^+\mu^-$ mass assignment.
      The large $\mathrm{K}\pi$ signal below $1.82\mass$ is
        kinematically separate from the region of interest.
      The distribution of events in the region above the \Dzero\ mass
is roughly flat.}
    \label{fig1}
  \end{center}
\end{figure}

The \dmm\ branching ratio, or upper limit, is determined using
\begin{equation}
  {\cal B}(\dmm) \le {\cal B}(\dpipi) \ 
\frac{N(\mu\mu)}{N(\pi\pi)}\ 
\frac{\epsilon(\pi\pi)}{\epsilon(\mu\mu)}\ 
\frac{a(\pi\pi)}{a(\mu\mu)},
\label{eq:limit}
\end{equation} 
where ${\cal B}(\dpipi) = (1.43\pm0.07)\times10^{-3}$ is the measured
normalization branching fraction~\cite{rpp}, 
$N(\mu\mu)$ and $N(\pi\pi)$ are the numbers of \dmm\ and \dpipi\
events observed, and
$\epsilon$ and $a$ are the efficiency and acceptance for each mode. 
Except for the requirement of muon identification, and the assignment
of different particle masses, the same selection
requirements are applied to both modes.
In this analysis, we determined the upper limit on the number of
signal events observed, $N(\mu\mu)$, by assuming that the number of events
found in the signal region is the sum of signal and background events, 
both obeying Poisson statistics.
Normalization was made to
\dpipi\ rather than the more numerous \dkpi\ decays.  Kinematically,
the \dpipi\ mode is nearly identical to \dmm, minimizing the
differences in acceptance and efficiency, and
introducing minimal systematic uncertainty to the result.  The
width of the reconstructed mass peak for two-body decays of the \Dzero\ in
\cdfii\ is about $10\dmass$, sufficient to separate \dkpi\
kinematically from \dpipi\ (Fig.~\ref{fig1}).

In the spirit of obtaining an unbiased result,
a ``blinded'' analysis was performed.
The data in the signal mass window were hidden
and the analysis cuts optimized without knowledge of their actual
impact on the result.  The optimization was performed on kinematically
similar but statistically independent events.  Only after all
selection criteria had been
fixed was the signal region ``unblinded'' and the final result determined.


We first outline the general event selection requirements common to
all the data samples used in the analysis and then discuss how they
are used to determine the quantities in Eq.~(\ref{eq:limit}).  All of
the samples consist of \dstdpi\ 
candidate decays coming from data sets
where all requisite detector components were functioning properly,
specifically, the SVX, COT, and CMU detectors, and the displaced-track
trigger chain.  
\Dzero\ candidates were formed from pairs of
oppositely charged ``trigger tracks'' that are ``CMU fiducial''.
Trigger tracks are tracks reconstructed offline that have been matched to
online SVT tracks.
A track that intercepts the active region of the
CMU when extrapolated from the COT through the magnetic field of the
detector is said to be CMU fiducial.

To select \Dzero\ candidates in a given decay mode, $K^+\pi^-$,
$\pi^+\pi^-$, or $\mu^+\mu^-$, we evaluated the invariant mass of each pair of
trigger tracks using the corresponding mass assignment and kept
candidates in the range $1.840\mass<M_{\mathrm{pair}}<1.884\mass$.
This corresponds to slightly more than $\pm2\sigma$ around the mean of the
\Dzero\ mass peak, $1.862\mass$.
The \Dstar\ tag
reduces non-\Dzero\ backgrounds and eliminates the 
mass peak in the $K\pi$ channel due to mis-assignment of the $K$
and $\pi$ masses.
$\Dstar^+\rightarrow\Dzero\pi^+_s$ decays 
were selected by combining an additional pion
track ($\pi_s$) with the \Dzero\ candidate and requiring the mass difference
$M_{\mathrm{pair}+\pi_s} - M_{\mathrm{pair}}$ to lie in the range
$144\dmass$ to $147\dmass$.  The $\pi_s$ track was not required to
be CMU fiducial or to be a trigger track, but it had to have the
Cabbibo-favored charge for the $K\pi$ decay.


The ratio
$\epsilon(\pi\pi)/\epsilon(\mu\mu)$ was determined from the muon
identification efficiency and the pion reconstruction efficiency,
measured in other analyses, as follows.  From a sample of
$\jpsi\rightarrow\mu^+\mu^-$ decays collected by a trigger requiring
one identified muon and one SVT track, the CMU identification
efficiency for the unbiased muon was measured offline as a function of
its transverse momentum $p_T$.  We convoluted the efficiency spectrum
with the $p_T$ spectrum of pions from \dpipi\ and determined the
effective dimuon identification efficiency to be
$\epsilon(\mu\mu)=0.800\pm0.030$.  Using a detailed GEANT~\cite{geant}
detector simulation, the pion reconstruction efficiency was found to
be $95\pm1\%$, yielding $\epsilon(\pi\pi)=0.90\pm0.02$, where the
inefficiency arises primarily from
hadronic interactions with detector material.
Combining these values we find $\epsilon(\pi\pi)/\epsilon(\mu\mu) =
1.13\pm0.04$.
Using the same detector simulation, we find the acceptance ratio
$a(\pi\pi)/a(\mu\mu) = 0.96\pm0.02$.


The number of \dpipi\ decays, $N(\pi\pi)$, was determined by fitting
the peak in the $\pi\pi$ invariant mass spectrum.
We performed a binned $\chi^2$ fit with Gaussian signal plus linear
background, as shown in Fig.~\ref{fig1}.
Both the mean and width of the Gaussian were free parameters
in the fit.
$N(\pi\pi)$ is the integral of the Gaussian over the $\pm22\dmass$ mass
window around $1.862\mass$.

The background to \dmm\ was taken as the sum of two contributions
having different mass spectra: 
a peaked contribution from \dpipi\ decays in which
both pions are misidentified as muons, and a relatively flat
background due to all other sources.  
The flat background
was estimated from the number of $\mu\mu$ candidates in a high mass
sideband spanning the range $1.90\mass<M_{\mu\mu}<2.05\mass$ with
both tracks identified as muons. 
Before muon identification of the tracks is required, the distribution
of events in the high mass sideband is found to be roughly constant, and
we assume that this remains true after requiring muon identification.
The expected flat
background is the number of sideband events scaled by the ratio of the
width of the signal region to the sideband region, $44/150$.

The misidentification background was estimated from the number of
\dpipi\ events reconstructed with the $\mu\mu$ mass assignment and lying 
in the $\pm22\dmass$ signal window 
(shaded area falling between the arrows in Fig.~\ref{fig1})
times the square of the probability for a pion to be misidentified as a muon.
The $\pi$-misidentification probability was determined from the sample
of \Dstar-tagged \dkpi\ events.
The average $\pi$-misidentification probability is $1.3\pm0.1\%$.

Three additional selection requirements were imposed.  To remove
instances in which the two \Dzero\ decay daughters extrapolate to the
same region of the CMU, potentially correlating the muon
identification of the two tracks, 
we cut on the azimuthal angle $\Delta\phi_{CMU}$ between their
projections into the CMU.
To suppress combinatoric backgrounds, we 
cut on the impact parameter with respect to the beamline, $d_{xy}$, of
the reconstructed \Dzero\ trajectory.
Further, we cut on the transverse decay length of the \Dzero\
candidate, $L_{xy}$.
The values of these cuts were optimized as described below.

We determined the optimal cut values by maximizing a figure of merit
given by $S/(1.5 + \sqrt{B})$ 
 \cite{Punzi}\ 
where $S$ and $B$
represent the number of signal and background events, respectively.
This quantity has desirable properties for an analysis where both the
signal and background are small:
it behaves as
$S/\sqrt{B}$ for large $B$ and it behaves as $S$ as the estimated
background approaches zero.
The constant in the denominator is chosen to 
favor cuts that maximize the discovery reach at $3\sigma$ significance.
To estimate $S$ in the optimization, we used the \dpipi\ sample.  To
estimate the misidentification component of $B$, we used a sample of
\dkpi\ decays in which both tracks were found to be
misidentified as muons.  To 
estimate all remaining contributions to $B$, we used the subset of the
high-mass $\pi\pi$ sideband sample in which one track was identified
as a muon and the other was not.
Note that the events used to estimate $B$ in the
optimization are distinct from the events used in the final background
estimate for the result.  The resulting selection requirements are:
 $|\Delta\phi_{CMU}| > 0.085\rm\,rad,$
 $|d_{xy}| < 150\rm\,\mu m,\ and$
 $L_{xy} < 0.45\rm\,cm.$
When applied to the samples used for optimization these cuts remove
approximately $58\%$ of the background events and $12\%$ of the
signal events.

Using the optimized selection requirements, $5.0\pm2.2$ events remain
in the high mass sideband, yielding $1.6\pm0.7$ expected 
from the flat component of the background.
The number of \dpipi\ events falling in the signal window when
reconstructed using the muon mass, multiplied by the square of the $1.3\%$
$\pi$-misidentification probability, yields
$0.22\pm0.02$ expected misidentification events.
The total expected background is $1.8\pm0.7$ events.
The number of events in the normalization mode is
$N(\pi\pi)=1412\pm54$ (Fig.~\ref{fig1}).
Using this background estimate and normalization, the 90\%\ confidence
level sensitivity~\cite{FandC} is $4.4\times10^{-6}$.

\begin{figure}[htb]
  \begin{center}
    \includegraphics[width=\columnwidth]{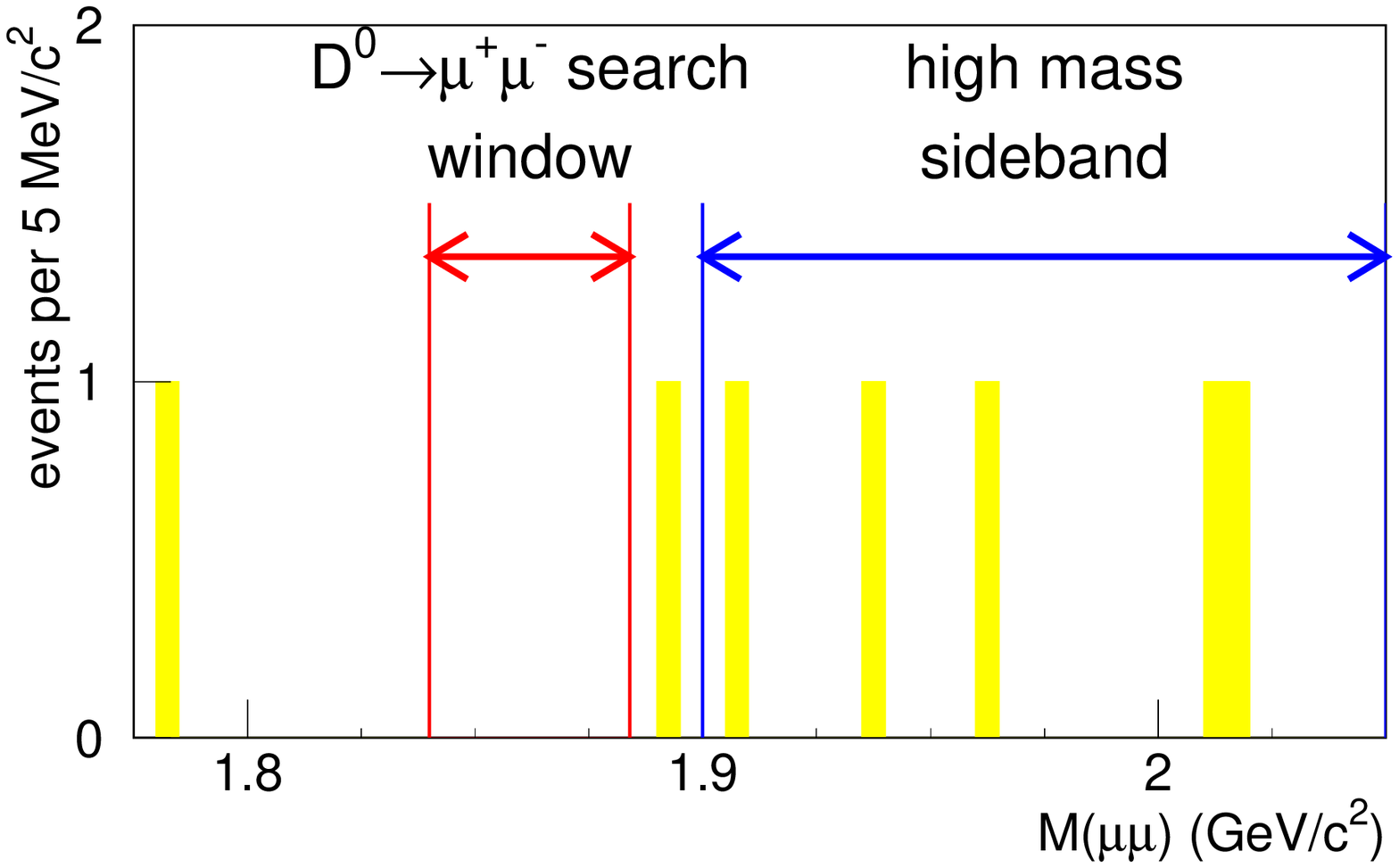}
    \caption{The mass distribution of candidate \dmm\ events.
  No events remain in the \Dzero\ mass region satisfying the
      event requirements.
      The events in the high mass sideband were used to estimate the
background from all sources other than misidentification of \dpipi.}
    \label{fig2}
  \end{center}
\end{figure}

We apply the optimized selection requirements to the signal region of
the $\mu\mu$ sample and find no events remaining, as displayed in
Fig.~\ref{fig2}.  
Conservatively taking the number of background events to equal zero,
the 90\% (95\%) confidence level upper limit on the
number of \dmm\ events is $2.3$ ($3.0$).  Using Eq.~(\ref{eq:limit})
we find an upper limit on the branching fraction of ${\cal
B}(\dmm)\le2.5\times10^{-6}$ ($3.3\times10^{-6}$) at the 90\%\ (95\%)
confidence level.

The uncertainties on $N(\pi\pi)$, $\epsilon(\pi\pi)/\epsilon(\mu\mu)$, 
$a(\pi\pi)/a(\mu\mu)$, and
${\cal B}(\dpipi)$ are incorporated into the limit using the
prescription of Cousins and Highland \cite{CandH}.
However, all of the uncertainties are smaller than 5\%\ and
have a negligible effect on the limit.


In summary, we have searched for the FCNC decay \dmm, using the new
displaced-track trigger of the \cdfii\ experiment.  This is the first
result from CDF in the field of rare charm decays.  To minimize bias
in the event selection, a blinded search was performed.  To minimize
dependence on Monte Carlo simulation, most of the needed quantities were
determined directly from the data.  No events were observed and we set
an upper limit on the branching ratio of
\begin{equation}
  \label{eq:result}
  {\cal B}(\dmm)\le2.5\times10^{-6} (3.3\times10^{-6})
\end{equation}
at the 90\%\ (95\%) confidence level.
This result improves on the best limits published to date.

\begin{acknowledgments} 
We thank the Fermilab staff and the technical staffs of the
participating institutions for their vital contributions.
This work was supported by
the U.S.~Department of Energy and National Science Foundation;
the Italian Istituto Nazionale di Fisica Nucleare;
the Ministry of Education, Culture, Sports, Science and Technology of
Japan;
the Natural Sciences and Engineering Research Council of Canada;
the National Science Council of the Republic of China;
the Swiss National Science Foundation;
the A.P.~Sloan Foundation;
the Bundesministerium fuer Bildung and Forschung, Germany;
the Korean Science and Engineering Foundation and the Korean Research
Foundation;
the Particle Physics and Astronomy Research Council and the Royal
Society, UK; the Russian Foundation for Basic Research; and the
Comision Interministerial de Ciencia y Tecnologia, Spain.
\end{acknowledgments}


\begin{thebibliography}{10}
\bibitem{chconj}
Throughout this paper inclusion of charge conjugate modes is implicit.
\end{thebibliography}
\end{document}